\def\rfr#1{eq. (\ref{#1})}

\def\bm#1{{\mbox{\boldmath$#1$\unboldmath}}}

\def\bar{\begin{eqnarray}}
\def\ear{\end{eqnarray}}
\def\bb{\bibitem}
\def\eqi{\begin{equation}}
\def\eqf{\end{equation}}
\def\eqia{\begin{eqnarray}}
\def\eqfa{\end{eqnarray}}
\def\rp#1#2{{#1\over#2}}
\def\ct#1{\cite{#1}}
\def\lb#1{\label{#1}}

\def\oc2{$\mathcal{O}(c^{-2})$}

\documentclass[11pt]{article}
\usepackage{amsmath,amsthm,amscd,amssymb}
\usepackage{latexsym}
\usepackage{graphicx,epsfig}

\begin{document}

\noindent{\bf \LARGE{Testing frame-dragging with the Mars Global
Surveyor spacecraft in the gravitational field of Mars}}
\\
\\
\\
{Lorenzo Iorio}\\
{\it Viale Unit$\grave{a}$ di Italia 68, 70125\\Bari (BA), Italy
\\tel. 0039 328 6128815
\\e-mail: lorenzo.iorio@libero.it}

\begin{abstract}
In this paper we propose an interpretation of the time
series of the RMS overlap differences of the out-of-plane part $N$
of the orbit of the Martian polar orbiter Mars Global Surveyor
(MGS) over a time span $\Delta P$ of about 5 years (14 November
1999-14 January 2005) in terms of the general relativistic
gravitomagnetic Lense-Thirring effect. It turns out that the average of such a time series over
$\Delta P$, normalized to the predicted Lense-Thirring
out-of-plane mean shift over the same time span, is $\mu=
1.0018\pm 0.0053$.
A linear fit of the considered time series over $\Delta P$  yields a normalized slope
$1.1\pm 0.3$ strengthening our interpretation of the determined MGS
out-of-plane features of motion as due to the gravitomagnetic
force of Mars.
\end{abstract}



\section{The gravitomagnetic Lense-Thirring effect}
In the framework of the weak-field and slow-motion linearized
approximation of general relativity \ct{Mas01}, the
gravitomagentic Lense-Thirring effect consists of tiny secular
precessions of the longitude of the ascending node $\Omega$ and
the argument of pericentre $\omega$ of the orbit of a test
particle freely moving in the gravitational field of a slowly
rotating body of mass $M$ and angular momentum $\bm J$ \ct{Len18,
Bar74, Cug78, Sof89, Ash93, Ior01}
\begin{equation}
\left\{
\begin{array}{lll}
\dot\Omega_{\rm LT}&=&\rp{2GJ}{c^2 a^3 (1-e^2)^{3/2}},\lb{leti}\\\\
\dot\omega_{\rm LT}&=&-\rp{6GJ\cos i}{c^2 a^3 (1-e^2)^{3/2}},
\end{array}
\right.
\end{equation}
where $G$ is the Newtonian gravitational constant, $c$ is the
speed of light in vacuum, $a, e$ and $i$ are the semimajor axis,
the eccentricity and the inclination of the orbital plane to the
central body's equator, respectively.  It maybe of interest to
recall that de Sitter in \ct{deS16} worked out, for the first
time, the gravitomagnetic pericentre precession in the case of
equatorial orbits ($i=0$ deg). In general, the gravitomagnetic
force affects the in-plane transverse and the out-of-plane
cross-track portions of the orbit of a test particle. Indeed, the
shifts along the radial $R$, transverse $T$ and normal\footnote{It
is customarily defined to be  positive along the orbital angular
momentum.} $N$ directions can be, in general, expressed in terms
of the Keplerian orbital elements as \ct{Chr88}
\begin{equation}
\left\{
\begin{array}{lll}
\Delta R &=&\sqrt{ (\Delta a)^2 + \rp{[ (e\Delta a + a\Delta e)^2 + (ae\Delta{\mathcal{M}})^2 ]}{2}},\\\\
\Delta T &=& a\sqrt{ 1+\rp{e^2}{2} }\left[\Delta{\mathcal{M}}
+\Delta\omega+\cos i\Delta\Omega +\sqrt{(\Delta
e)^2+(e\Delta{\mathcal{M}})^2} \right],\\\\
\Delta N &=& a\sqrt{ \left(1+\rp{e^2}{2}\right)\left[\rp{(\Delta i
)^2}{2}+(\sin i\Delta\Omega)^2\right] },
\end{array}
\right.
\end{equation}
where $\mathcal{M}$ is the mean anomaly. The Lense-Thirring shifts
are, thus
\begin{equation}
\left\{
\begin{array}{lll}
\Delta R_{\rm LT} &=& 0,\\\\
\Delta T_{\rm LT} &=& a\sqrt{ 1+\rp{e^2}{2}
}\left(\Delta\omega_{\rm LT}+\cos i\Delta\Omega_{\rm LT}\right),\\\\
\Delta N_{\rm LT} &=& a\sqrt{1+\rp{e^2}{2}}\sin i\Delta\Omega_{\rm
LT}.\lb{chris}
\end{array}
\right.
\end{equation}
For an equatorial orbit there is only the transverse shift $\Delta
T_{\rm LT} $, while for a polar ($i=90$ deg) satellite only the
out-of-plane shift $\Delta N_{\rm LT}$ is present. This fact can
be more intuitively grasped in the following way. In general, a
time-variation of the argument of pericentre $\omega$, which is an
angle entirely residing in the orbital plane delimited by the line
of the nodes and the apsidal line, can be obtained by subtracting
the projection of the nodal motion to the orbital plane, i.e.
$\cos i\dot\Omega$, from the precession of the longitude of the
pericentre $\varpi$. For a polar orbital geometry the projection
of the node's motion to the orbital plane is zero. According to
the well-known Gauss variation equations \ct{Roy03}, the remaining
shift induced by a small perturbing acceleration $\bm A$ is due to
its in-plane radial $A_R$ and transverse $A_T$ components. The
Lense-Thirring effect is due to a Lorentz-like acceleration
\ct{Lic06} \eqi\bm A_{\rm LT}=-2\left(\rp{\bm v}{c}\right)\times
\bm B_g;\lb{fgm}\eqf the gravitomagnetic field \ct{Lic06} \eqi\bm
B_g=\rp{G[3\bm r(\bm r\cdot \bm J )-r^2 \bm
J]}{cr^5}\lb{gmfield}\eqf for $i=90$ deg lies in the test
particle's orbital plane which contains also $\bm J$, so that the
net gravitomagnetic force is entirely normal to the orbital plane.

Up to now, many attempts to test gravitomagnetism
were implemented in the weak-field and slow-motion scenarios of
the gravitational fields of some Solar System's bodies. Lense and
Thirring  considered the natural satellites of Jupiter and Saturn \ct{Len18}, but the
possibility of using the jovian galilean moons is not yet viable \ct{Lai05}. In regard to the Sun and
its rocky planets \ct{Ior05c}, the predictions of
general relativity for the Lense-Thirring perihelia precessions  were, in fact, found in agreement with the estimated corrections to the
secular rates of the perihelia of the inner planets
\ct{Pit05}, but the errors are still large.
As far as the Earth is concerned,  a test of the Lense-Thirring
effect on the orbit of a test particle was reported in \ct{Ciu04}.
Their authors analyzed the data of the laser-ranged LAGEOS and
LAGEOS II satellites  by using an observable independently
proposed in \ct{Rie03a, Rie03b} and \ct{Ior04} involving the nodes of the LAGEOS satellites. The idea of using
the LAGEOS satellite and, more generally, the Satellite Laser
Ranging (SLR) technique to measure the Lense-Thirring effect in
the terrestrial gravitational field with the existing artificial
satellites was put forth for the first time by Cugusi and
Proverbio in 1977-1978 \ct{Cug77, Cug78}; tests actually started in 1996 with LAGEOS and LAGEOS II \ct{Ciu96}.
Although the outcome of the test in
\ct{Ciu04} is in agreement with the predictions of general
relativity, the realistic evaluation of the obtained total
accuracy raised a debate \ct{Ior05a, Ior05b, Ior06a, Ior06b,
Ciu05, Luc05}. The GP-B spacecraft \ct{Eve74, Eve01} was launched
in April 2004 with the aim of measuring another gravitomagnetic
effect, i.e. the Schiff precession \ct{Pug59, Sch60}  of the spins
of four superconducting gyroscopes carried onboard. Preliminary results, released in April 2007,
yield a measurement uncertainty of the order of $256-128\%$; the final results
should be announced within the end of 2007, with a goal of an about $13\%$ accuracy.
(http://einstein.stanford.edu/).
Recently, a preliminary $6\%$ LT test on the orbit of the Mars
Global Surveyor (MGS) spacecraft (see Table \ref{tavolaMGS} for
its orbital parameters) in the gravitational field of Mars was
reported in \ct{Ior06c}; indeed, the predictions of general
relativity are able to accommodate, on average, about $94\%$ of
the Root-Mean-Square (RMS) overlap differences in the out-of-plane part of the MGS orbit
over 5 years.
{\small\begin{table}\caption{ Orbital parameters of MGS
\ct{Kon06}; $a$ is the semimajor axis, $i$ is the inclination to the Mars'equator and
$e$ is the eccentricity. }\label{tavolaMGS}
\begin{center}
\begin{tabular}{ccc}\noalign{\hrule height 1.5pt}
\hline $a$ (km) & $i$ (deg) & $e$ \\
\hline $3796$  & $92.86$ & $0.0085$\\
\hline
\noalign{\hrule height 1.5pt}
\end{tabular}
\end{center}
\end{table}}
Finally, it must be mentioned that, according to Nordtvedt
\ct{Nor88a, Nor03}, we would already have evidences of the
existence of the gravitomagnetic interaction. Indeed, the
multi-decade analysis of the Moon'orbit by means of the Lunar
Laser Ranging (LLR) technique yields a comprehensive test of the
various parts of order $\mathcal{O}(c^{-2})$ of the post-Newtonian
equation of motion. The existence of gravitomagnetism as predicted
by general relativity would, thus, be inferred with high precision
from the reconstruction of certain periodic terms in the lunar
motion. A $0.1\%$ accuracy is claimed in \ct{Mur07}, but,
according to Kopeikin \ct{Kop07}, LLR would not be capable to
detect gravitomagnetic effects; Murphy et al. replied in \ct{MurR07}. 
Other comments on \ct{Mur07} are in \ct{Ciu07}.
Also the radial motion of the
LAGEOS satellite would yield another confirmation of the existence
of the gravitomagnetic interaction \ct{Nor88b}.

In this paper we perform a  test of the Lense-Thirring effect by
improving the accuracy of the previous one with MGS  \ct{Ior06c}
by means of a different data set, a refinement of the overall
calculation and a realistic and conservative assessment of the
error estimate. The details of the adopted strategy can be found
in \ct{Ior06c, Kon06}; here we will briefly recall them.

\section{Using the MGS data in the gravitational field of Mars}
In \ct{Kon06} six years of MGS Doppler and range tracking data and
three years of Mars Odyssey Doppler and range tracking data were
analyzed in order to obtain information about several features of
the Mars gravity field summarized in the global solution MGS95J.
As a by-product, also the orbit of MGS was determined with great
accuracy. In Figure 3 of \ct{Kon06} the time series of the RMS
overlap differences of the radial, transverse  and normal portions
of its orbit over the entire mission (4 February 1999$-$14 January
2005) are shown; their averages over such a time span amount to
\ct{Kon06}
\begin{equation}
\left\{
\begin{array}{lll}
\left\langle\Delta R_{\rm diff}\right\rangle = 15\ {\rm cm},\\\\
\left\langle\Delta T_{\rm diff}\right\rangle = 1.5\ {\rm m},\\\\
\left\langle\Delta N_{\rm diff}\right\rangle= 1.6\ {\rm m}.
\end{array}\lb{ressi}
\right.
\end{equation}
The RMS overlap differences are customarily used in satellite
geodesy as indicators of the overall orbit accuracy \ct{Tap04},
concisely accounting not only for the measurement errors, but also
for all the systematic bias due to the mismodelled/unmodelled
forces acting on a spacecraft \ct{LucBal06}. Errors common to adjacent arcs are, by construction, cancelled;
it is not so for effects growing in time like the one we are interested in here. Thus, any  analytical calculation
of possible systematic errors must ultimately cope with that is
actually shown by the RMS overlap differences. It is important to note that
in the suite of dynamical force models used to reduce the gathered
observations general relativity is, in fact, present, but only
with the gravitoelectric Schwarzschild-like part: the
gravitomagnetic one is not included, so that, in general, it is
fully accounted for by the RMS orbit overlap differences. From
\rfr{chris} it turns out that only the out-of-plane portion of the
orbit of a polar satellite like MGS is affected by
gravitomagnetism via the Lense-Thirring effect: from \rfr{leti}
and \rfr{chris} the mean gravitomagnetic normal shift, averaged
over a temporal interval $\Delta P$, is\footnote{A 1/2 factor appears as a result of the time-average of a linear signal like the gravitomagnetic one over a time interval $\Delta P$.} \eqi \left\langle\Delta
N_{\rm LT}\right\rangle = \rp{GJ\Delta P\sin i\sqrt{ 1+\rp{e^2}{2}
}}{c^2 a^2(1-e^2)^{3/2}}.\lb{ltshift}\eqf Moreover, no empirical
out-of-plane accelerations were fitted in the data reduction
process; this was only done for the along-track component of MGS
orbit to account for other non-gravitational forces like thermal,
solar radiation pressure and drag mismodelling.
Thus, it is unlikely that the out-of-plane Lense-Thirring signal
was removed from the data in the least-square procedure. Moreover,
if the relativistic signature was removed so that the determined
out-of-plane RMS overlap differences were only (or mainly) due to other causes
like mismodelling or unmodelling in the non-gravitational forces,
it is difficult to understand why the along-track RMS overlap differences
(middle panel of Figure 3 of \ct{Kon06}) have almost the same
magnitude, since the along-track component of the MGS orbit is
much more affected by the non-gravitational accelerations (e.g.
the atmospheric drag) than the out-of-plane one. Finally, the following consideration should further
enforce our conclusion. The improved modelling of the non-gravitational forces acting on MGS introduced in \ct{Kon06}
with respect to
previous works \ct{Lem01, Yua01} in which the Lense-Thirring effect was not modelled as well,
only affected in a relevant way the along-track RMS overlap differences (a factor 10 better than \ct{Lem01, Yua01}),
not the normal ones (just a factor 2 better than \ct{Lem01, Yua01}).

By using a time span $\Delta P$ of slightly more than 5 years (14
November 1999$-$14 January 2005), so to remove the first months of
the mission more affected by orbital maneuvers, high gain antenna deployment and
non-gravitational perturbations, it is possible to obtain \eqi
\left\langle\Delta N_{\rm diff}\right\rangle = 1.613\ {\rm
m}.\lb{meashift} \eqf Let us, now, compare the result of
\rfr{meashift} with the mean gravitomagnetic shift predicted by
\rfr{ltshift} over the same time span; according to our
hypothesis, the Lense-Thirring force of Mars is responsible for
the averaged normal signature of \rfr{meashift}. The
figures of Table \ref{tavolaMars} for the Mars'relevant parameters
and those of Table \ref{numval} for the fundamental constants
entering the calculation yield \eqi \mu_{\rm meas}\equiv
\left.\rp{\left\langle\Delta N_{\rm
diff}\right\rangle}{\left\langle\Delta N_{\rm
LT}\right\rangle}\right|_{\rm \Delta P}= 1.0018\pm
0.0053;\lb{measu}\eqf according to our hypothesis, $\mu_{\rm
expected}=1$.
{\small\begin{table}\caption{Relevant parameters of Mars used in
the text. The parameter $\alpha\equiv C/M{\mathcal{R}}^2$ is the
normalized polar moment of inertia, ${\mathcal{R}}$ is the
reference Mars radius in MGS95J model, while $\mathcal{P}$ is the
Martian sidereal period of rotation. From such values and from
$J=\alpha M{\mathcal{R}}^2 2\pi/{\mathcal{P}}$ it can be obtained
$J=(1.92\pm 0.01)\times 10^{32}$ kg m$^2$ s$^{-1}$ for the Martian
angular momentum.}\label{tavolaMars}
\begin{center}
\begin{tabular}{cccc}\noalign{\hrule height 1.5pt}
\hline $GM$ ($10^9$ m$^3$ s$^{-2}$) & $\mathcal{R}$ (km) & $\alpha$ & ${\mathcal{P}}$ (d)\\
\hline $42828.37440\pm 0.00028$ \ct{Kon06} & $3396.0$ \ct{Kon06} & $0.3654\pm 0.0008$ \ct{Kon06}& $1.02595675$ \ct{Sei92}\\
\hline
\noalign{\hrule height 1.5pt}
\end{tabular}
\end{center}
\end{table}}
\begin{table}
\caption{ Values used for the defining, primary and derived
constants
(http://ssd.jpl.nasa.gov/?constants$\#$ref).}\label{numval}

\begin{tabular}{cccccc} \noalign{\hrule height 1.5pt}
constant & numerical value & units & reference\\
\hline $c$ & 299792458 & m s$^{-1}$ & \ct{Moh05}\\
$G$ & $(6.6742\pm 0.0010)\times 10^{-11}$ & kg$^{-1}$ m$^3$ s$^{-2}$ & \ct{Moh05}\\
$1$ mean sidereal day & $86164.09054$ & s & \ct{Sta95}\\
$1$ mean sidereal year & $365.25636$ & d & \ct{Sta95}\\
\hline

\noalign{\hrule height 1.5pt}
\end{tabular}

\end{table}
The error was conservatively evaluated as
\eqi\delta\mu\leq\delta\mu|_{J} + \delta\mu|_{G} +
\delta\mu|_{a},\eqf with
\begin{equation}
\left\{
\begin{array}{lll}
\delta\mu|_J\leq \mu\left(\rp{\delta J}{J}\right) = 0.0052,\\\\
\delta\mu|_G\leq \mu\left(\rp{\delta G}{G}\right) = 0.0001,\\\\
\delta\mu|_a\leq 2\mu\left(\rp{\delta a}{a}\right) = {\mathcal{O}(10^{-8})}.
\end{array}\lb{errsym}
\right.
\end{equation}
As can be noted, the major source of bias is represented by the
angular momentum of Mars. Its uncertainty was evaluated
as\eqi\delta J\leq \delta J|_{\mathcal{R}} + \delta J|_{\alpha} +
\delta J|_M,\eqf with
\begin{equation}
\left\{
\begin{array}{lll}
\delta J|_{\mathcal{R}}\leq
2J\left(\rp{\delta{\mathcal{R}}}{{\mathcal{R}}}\right)= 7\times 10^{29}\ {\rm kg} \ {\rm m}^2\ {\rm s}^{-1},\\\\
\delta J|_{\alpha} \leq  J\left(\rp{\delta\alpha}{\alpha}\right) = 4\times 10^{29}\ {\rm kg} \ {\rm m}^2\ {\rm s}^{-1},\\\\
\delta J_M \leq J\left[\rp{\delta(GM)}{GM}\right] = {\mathcal{O}(10^{24})}\
{\rm kg} \ {\rm m}^2\ {\rm s}^{-1}.
\end{array}\lb{ERRJ}
\right.
\end{equation}
We linearly added all the errors in a conservative way; in regard
to $\alpha$ and $GM$ this is particularly important because they
are fit-for parameters of the global solution and, consequently,
are certainly affected by correlations which were not explicitly
released in \ct{Kon06}. It must also be noted that the
uncertainties quoted in Table \ref{tavolaMars} are realistic, not
the mere formal, statistical errors; indeed, the quoted
uncertainty in $\alpha$ is, e.g., 15 times the formal one
\ct{Kon06}. Also the formal error in $GM$ was re-scaled
\ct{Kon06}. We also considered an uncertainty in ${\mathcal{R}}$
by assuming as a realistic measure of it the difference
$\delta{\mathcal{R}}=6080$ m between the reference MGS95J value
for the equatorial radius quoted in Table \ref{tavolaMars} from
\ct{Kon06} and the value  3389.92 km released in \ct{Yod95},
instead of the error of 0.04 km quoted in \ct{Yod95}. In regard to
the error in $a$, we assumed from \rfr{ressi} $\delta a\approx 15$
cm in view of the very small eccentricity of the MGS orbit.


A further confirmation of the validity of our interpretation of
the MGS out-of-plane motion in terms of the gravitomagnetic force
of Mars comes from a linear\footnote{It may be just the case to
note that, since the plots in Fig. 3 of \ct{Kon06} are
semi-logarithmic, one should not visually look for a straight line
in them.} fit of the time series of the RMS overlap differences
over $\Delta P$. Indeed, the estimated slope is $-0.69\pm 0.20$ m
yr$^{-1}$, while general relativity predicts a Lense-Thirring
out-of-plane rate of 0.62 m yr$^{-1}$, with the normal direction
defined positive along the orbital angular momentum. The minus
sign of the fit is due to the fact that in the analysis of
\ct{Kon06} the MGS normal direction is, instead, defined positive
in the opposite direction (Konopliv A, private communication
2007). Incidentally, let us note that should we assume such a fit
as indicator of the existence of the Lense-Thirring effect, the predictions of general relativity
would be confirmed within the experimental error which, however,
would be larger than the one obtained by taking the mean of the
time series of $N$ over $\Delta P$.

Another important remark concerning our measurement is the
following one: the interpretation of the MGS orbital features presented
in this paper and their production from the collected data were
made by distinct and independent subjects in different times;
moreover, the authors of \ct{Kon06}  had not the Lense-Thirring
effect in mind at all when they performed their analysis. Thus,
the possibility of any sort of (un)conscious gimmicks, tricks or
tweaking the data because the expected answer was known in advance
\ct{Wei93} is a priori excluded in this test.
\section{Conclusions}
We conclude by summarizing our main result: the ratio $\mu$ of the
mean of the time series of the cross-track $N$ RMS overlap differences
of MGS from 14 November 1999 to 14 January 2005 to the mean of the
Lense-Thirring shift over the same time interval amounts to
$\mu=1.0018\pm 0.0053$. Our interpretation of the features of the
out-of-plane motion of MGS in terms of the gravitomagnetic force
of Mars is enforced by a linear fit of the $N$ time series,
although the error would be, in this case, larger.

It must be noted that the present analysis is based on the so far
processed data in \ct{Kon06}; other approaches which could
fruitfully be used in the future may consist in re-processing the
entire MGS data set including in the dynamical force models the
gravitomagnetic force as well and checking if the average of the $N$
time series gets smaller, and/or including
a further solve-for parameter in the global least-squares solution
accounting for the Lense-Thirring effect itself.

Finally, let us note that, although we lost contacts with MGS
since early November 2006, more independent tests could be, in principle, performed
in the near future when a sufficient amount of data from the
Odyssey and, especially, Mars Reconnaissance Orbiter (MRO)
orbiters will be available and processed.

\section*{Acknowledgments}
I gratefully thank A. Konopliv, NASA Jet Propulsion Laboratory
(JPL), for having kindly provided me with the entire MGS data set.
I am also grateful to the anonymous referees for their useful
comments and suggestions. Finally, I am indebted  to E. Lisi, P.
Colangelo, G. Fogli, S. Stramaglia, and N. Cufaro-Petroni (INFN,
Bari) for very stimulating discussions.


\end{document}